\documentclass{article}
\PassOptionsToPackage{round, comma}{natbib}
\usepackage[preprint]{arxiv}
\usepackage[dvipsnames]{xcolor}
\usepackage[utf8]{inputenc} 
\usepackage[T1]{fontenc}    
\usepackage{hyperref}       
\hypersetup{
    colorlinks=true,
    linkcolor=black,
    citecolor=CadetBlue
}
\usepackage{url}            
\usepackage{booktabs}       
\usepackage{amsfonts}       
\usepackage{nicefrac}       
\usepackage{microtype}      
\usepackage{graphicx}
\usepackage{amsmath, amssymb, amsthm}
\usepackage{parskip}
\usepackage{flushend}
\usepackage{caption} 
\usepackage{subcaption}
\usepackage{array,color,stfloats,chngpage,times,datetime,epstopdf}
\usepackage{lineno}
\usepackage[onehalfspacing]{setspace}
\usepackage{algorithm} 
\usepackage[noend]{algpseudocode}
\makeatletter
\def\BState{\State\hskip-\ALG@thistlm}
\makeatother

\title{Bayesian Generative Models for Knowledge Transfer in MRI Semantic Segmentation Problems}

\author{
  Anna Kuzina  \\
   ADASE\\
   Skoltech\\
   Moscow, Russia \\
  \texttt{a.kuzina@skoltech.ru} \\
   \And
 Evgenii Egorov  \\
   ADASE\\
   Skoltech\\
   Moscow, Russia \\
  \texttt{e.egorov@skoltech.ru} \\
  \And
 Evgeny Burnaev \\
   ADASE\\
   Skoltech\\
   Moscow, Russia \\
  \texttt{e.burnaev@skoltech.ru} \\
}

\begin{document}
\maketitle

\begin{abstract}
Automatic segmentation methods based on deep learning have recently demonstrated state-of-the-art performance, outperforming the ordinary methods. Nevertheless, these methods are inapplicable for small datasets, which are very common in medical problems. To this end, we propose a knowledge transfer method between diseases via the Generative Bayesian Prior network. Our approach is compared to a pre-train approach and random initialization and obtains the best results in terms of Dice Similarity Coefficient metric for the small subsets of the Brain Tumor Segmentation 2018 database (BRATS2018).
\end{abstract}


\section{Introduction}
Magnetic resonance imaging (MRI) is a medical imaging technique used in radiology to form pictures of the anatomy of some part of the human body. It is used as a diagnostic tool for various types of cancer, diseases of the central nervous system, such as multiple sclerosis or epilepsy \citep{hammers2007automatic, Pipeline2018, Epilepsy2018}, depression \citep{sheline20003d, DepressionAWE2018} and in plenty other cases \citep{cciccek20163d, ronneberger2015u}. Recent advances in computer vision revealed a high potential for application of neural networks in the medical problems: classification of MRI or CT for disease diagnosis, automatic detection and segmentation of different pathologies \citep{davatzikos2008individual, gong2007classification, 3DANNMRI2018}. Even though it is unlikely that these models will be used as a diagnostic tool without any human intervention in the nearest future, they could be beneficial serving as decision support systems.

Semantic segmentation of MRI scans is an essential but highly challenging task.  Accurate segmentation can simplify and speed up the work of radiologist, reduce the risk of mistakes by automatic detection of tumors \citep{kohl_adversarial_2017}, multiple sclerosis plaques \citep{rey_automatic_2002}, hemorrhages \citep{davuluri_hemorrhage_2012, guerrero_white_2017} or other disease manifestations \citep{wachinger_deepnat:_2018}. It is also applicable for analysis and quantification of some illnesses. For example, currently, the exact volume of affected brain areas of patients with multiple sclerosis is not calculated due to the extreme difficulty of this task. Instead, a very rough approximation is used while exact information about affected volumes in practice may be highly useful for understanding the progression of the disease. 

State-of-the-art methods for semantic segmentation imply the use of deep neural networks, which usually have millions of tuning parameters, hence demanding a large amount of labelled training samples to avoid overfitting. At the same time, manual labelling of the MRI with tumors or other manifestation of the disease, is time consuming and expensive. Consequently, in most cases only tiny datasets are available for training. As a result, methods which need less labelled examples for training are of great significance. To this end, we can exploit knowledge from existing labelled datasets. 

Medical imaging dataset has several crucial peculiarities, which one should take into account while solving semantic segmentation problem with the small training dataset. We can group them into image preprocessing, prediction postprocessing, selection of network architecture and specificity of the transfer learning from data with a different disease. Preprocessing includes image alignment, skull-stripping, normalization of the images within a given dataset \citep{litjens2017survey}. A variety of MRI protocols are available with or without the use of contrast agents. These protocols allow the setting up of different contrasts among the various tissues within the same organ system. Thus, the quality of the segmentation heavily depends on this feature of the dataset.

Depending on the dataset, different postprocessing of the obtained prediction may be required. For example, it is a common problem, that the full 3D scan does not fit into memory, and one has to use patches to obtain predictions. Predictions for overlapping patches are further combined by giving a higher weight to the pixels in the centre since they are known to produce better predictions. Moreover, for some problems, it is known that predicted mask could not contain more that one connected component, e.g. when a separate organ or it's part is being segmented. In this case, postprocessing could also remove all the extra prediction, which may drastically boost the performance.

Furthermore, the choice of network architecture is a crucial step. Semantic segmentation problem is usually solved in computer vision by fully convolutional networks with architectures similar to U-Net  \citep{ronneberger2015u}.  U-Net with 3D convolutions also known as V-Net \citep{milletari_v-net:_2016} is extensively applied to various types of medical images \citep{deniz_segmentation_2018, livne_u-net_2019, guerrero_white_2017, milletari_v-net:_2016, ronneberger2015u}. The state-of-the-art approaches consider additional regularization with training multitarget networks and also the ensembling of the models \citep{myronenko_3d_2018} or cascade models by stacking several V-Nets \citep{isensee2018nnu}. 

Finally, there is a common practice to apply transfer learning techniques, when the size of the target training datasets is not sufficient. There exist several large publicly available dataset with labelled segmentation, which may be used to transfer knowledge to smaller ones. Nevertheless, these images may be pretty different in terms of diseases, modality, protocols and preprocessing methods, which leads to extra difficulties. In this work, we address the problem of knowledge transfer between medical datasets when source dataset potentially contains relevant information for the given problem (e.g. it depicts scans of the same organ), but still comes from the different domain, complicating the work of the conventional transfer learning techniques.

\subsection{Transfer Learning approach} 
Transfer learning is a set of techniques from machine learning, used to store knowledge from one problem or dataset and apply it to another but similar problem \citep{pan2010survey}. In deep learning, it is usually performed by network initialization with weights trained on source dataset and fine-tuning on a target dataset. If the size of the target dataset is too small, some parameters of the network may be frozen to avoid overfitting. This approach can be beneficial for the segmentation of medical images \citep{havaei2016deep}, but the degree to which it will be useful highly depends on the source and target datasets similarity. \cite{van2015transfer} applied transfer learning to support vector machine classifier in the setting, where the source and target datasets only differ in scanners and acquisition protocols. The authors showed that with a small target dataset transfer learning considerably outperforms common supervised learning approach. \cite{ghafoorian2017transfer} were also using very similar datasets for transfer learning in white matter hyperintensities segmentation problem and obtained higher dice similarity coefficient when the model was trained on the target and fine-tuned on the source domain. The authors of both papers assumed that source dataset is almost the same as a target one with only small differences, such as scanner type or voxel size to be present.

 \citet{margeta2017fine} used fine-tuning to solve classification task on MRI scans. A convolutional neural network was pre-trained on a dataset with natural images, which is somewhat irrelevant for their problem and therefore requires fine-tuning of the whole model with a relatively big dataset of 215 MRI scans. \cite{zhou2017fine} proposed using continuous fine-tuning when training dataset is steadily expanded with images, labelled by the current version of the model. The authors suggested starting from the pre-trained network and choosing the most confident predictions of the model to include them into the training set. The main restriction of this approach is the fact that the method requires unlabelled data from the same domain. Moreover, the authors suggested working only with patches of images to assess the confidence of the algorithm, which might be less practical for tasks different from classification, such as detection or segmentation. \citet{han2018deep} exploited network pre-trained on a large number of X‐ray computed tomography (CT) to restore high-resolution MRI from under-sampled k-space data with few training MR observations available. 

\citet{christodoulidis2017multisource} showed improvement in lung tissue pattern classification accuracy when fine-tuning the model trained on six open-source texture databases separately and taking an ensemble of all these models. The authors determined that transfer learning from a single dataset does not provide a stable increase in accuracy and sometimes even performs worse than random initialization. \cite{li2018novel} proposed a novel approach, which helps to transfer knowledge from healthy subjects to new disease classification problem. They showed improvement in accuracy, sensitivity, specificity over deep neural network trained from scratch. But the authors only use fully connected layers, working with features exctracted from the functional MRI, rather that with raw images.

Another branch of work suggests dealing with smaller sample size, using mixed supervision models \citep{mlynarski2018deep, shah2018ms}. These papers highlight that for medical image segmentation we cannot rely on transfer learning of parameters from networks (pre-)trained for analysis of natural images. Hence, the authors proposed to simultaneously use high-quality expensive labelling with lower-quality but cheap labels for training (mixed supervision). Despite of the interesting results, the developed method considers the case of partially available expensive labels from one dataset. However, we consider the case of different datasets with different diseases.  

Transfer learning may be also considered as a special case of the domain adaptation problem \citep{wilson2018adversarial}, when one aims to take model trained on one domain (referred to as source) and adapt it to perform just as well on a new target domain. 

Finally, \cite{elsayed2018adversarial} suggest novel approach, adversarial reprogramming of the neural networks. The paper considers an additive perturbation to the network input to apply the adversarial reprogramming. The authors demonstrated adversarial reprogramming on classification tasks in the  2D image domain (MNIST classification, and CIFAR-10 classification). To apply the approach one should define a hard-coded mapping function from source labels to the adversarial task labels. Therefore we can not apply this approach to the segmentation tasks. Moreover, the method is applicable only for datasets with images of smaller spatial size than that of the source dataset. Hence, it is an interesting research problem to adapt the proposed technique for segmentation tasks of 3D MR images, however, it is out of the scope of our paper.  

 In this paper, we propose a method for knowledge transfer between diverse neuroimaging datasets. Conceptually, our approach consists of the following steps: we solve the semantic segmentation problem for a small labelled training dataset. Provided a larger dataset, referred to as the source, which may differ from the target dataset drastically in terms of the modality, resolution or other properties. Proposed method outperform straightforward fine-tuning on studied semantic segmentation problem.

When dealing with a small dataset along with the multidimensional model, there exists a high risk of overfitting. Experiments show that filters of different segmentation networks often exhibit similar structure, which could be exploited for regularization purposes. Probabilistic formulation of the model allows us to apply these restrictions on the weights formally using the method described below.

At the first stage, the source dataset is used to train a segmentation network.  Following the assumption that kernels from this model have a useful structure for the target segmentation problem, generative model --- Variational Autoencoder (VAE) \citep{kingma_vae} is trained on the weights from the source network which tries to approximate the distribution of the kernels. Finally, to solve the target problem, we fit the segmentation network with the same architecture as in the first point but with the generative model used as a prior distribution over the weights.

The rest of the paper is organized as follows: in Sections \ref{sec:unet} - \ref{sec:dwp} we discuss U-Net architecture \citep{ronneberger2015u}, which was used for semantic segmentation, describe deep Bayesian approach for training neural networks with prior distribution over parameters and, finally, explain how we can learn prior distribution from data and apply it to variational inference to perform knowledge transfer. Section \ref{sec:dataset} is devoted to the medical datasets, that were used for the experiments, in Sections \ref{sec:eval}- \ref{sec:exp} more practical details, such as metrics, loss functions and experimental setup are presented. Section \ref{sec:result} discusses the results of the experiments, where we compare the proposed approach with random initialization and pre-trained weight initialization. Finally, in Section \ref{sec:discussion} we discuss the key findings of the study, potential drawbacks and outline for the future work.
\section{Material and methods} \label{sec:main}

In this part, we shall discuss U-Net architecture, which serves as a foundation for all the experiments in this work. Then we discuss the approximate Bayesian approach, stochastic variational  inference \citep{hoffman2013stochastic} in deep neural networks and the importance of prior distribution selection. This part is crucial for the understanding of Deep Weight Prior (DWP) \citep{atanov_deep_2018}, which allows us to transfer knowledge among datasets. The idea of  DWP lies in the fact that we learn the prior distribution of convolutional filters in the form of a generative model, instead of using parametric distribution. Since we get kernels from the network trained on source dataset to learn the prior and further exploit it for variational inference on the target dataset, this approach can be considered as a transfer learning technique. Finally, we proceed to the description of the practical part, including datasets, validation methods, loss function and complete experiment setup, which evaluates the performance of the proposed approach.

\subsection{3D U-Net}  \label{sec:unet}
\begin{figure}
\centering
\includegraphics[width=15cm]{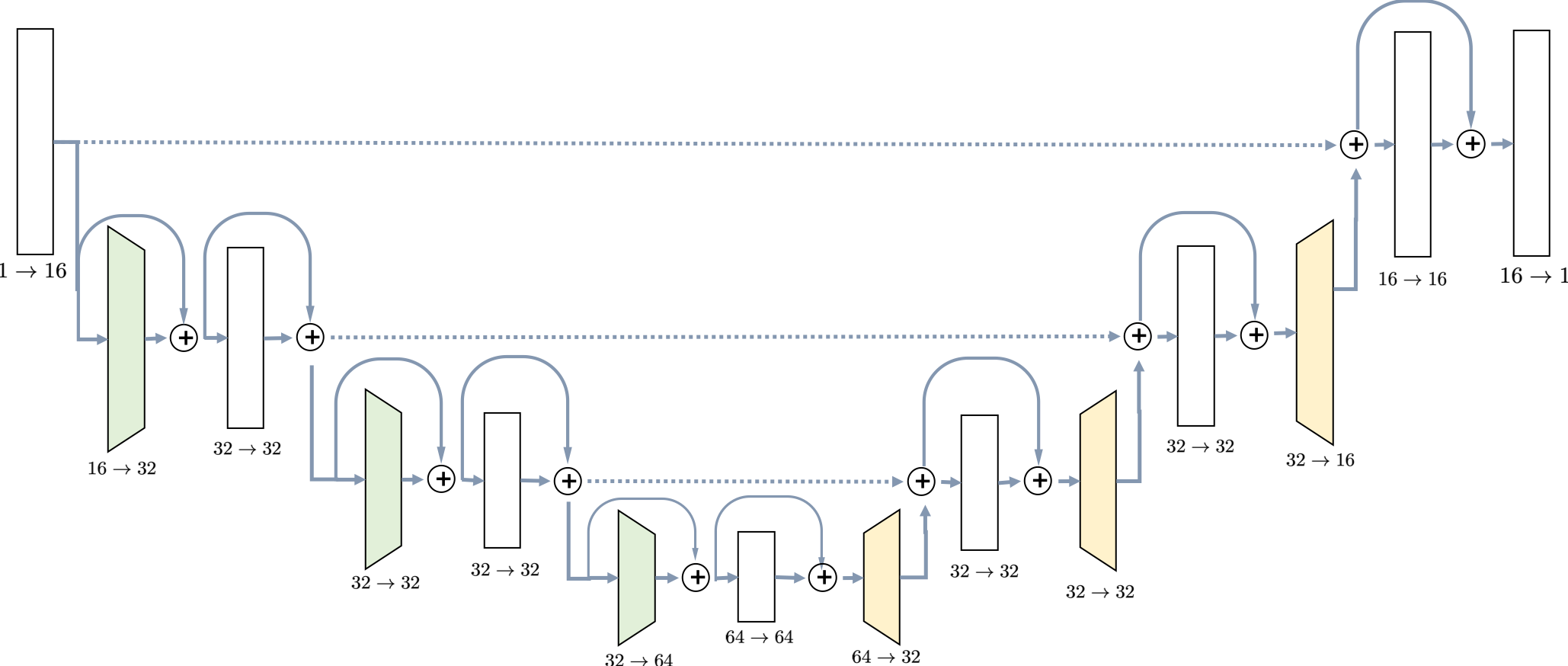}
\caption{U-Net architecture used in the experiments contains ResNet-like block both in Encoder and Decoder parts with skip-connections.}
\label{fig:unet}
\end{figure}

U-Net \citep{ronneberger2015u} was chosen due to its popularity  and experimentally proven efficiency for MRI semantic segmentation tasks \citep{deniz_segmentation_2018, livne_u-net_2019, guerrero_white_2017, milletari_v-net:_2016}. The detailed architecture of the network is shown in Figure \ref{fig:unet}. It consists of downsampling blocks, coloured in green, upsampling bocks (yellow) and simple blocks which do not change spatial resolution of the image. The chosen architecture has 726480 parameters, estimated from a training set of $20$ or $10$ images. Since U-Net is a fully convolutional network, the number of parameters does not depend on the input size. Regardless of the initial resolution, each input is compressed by the factor of 8 in the encoder part of the network and upsampled back to the initial size in the decoder. For instance, BRATS18 \citep{menze_multimodal_2015} which is initially cropped to $[152, 184, 144]$ pixels, gets compressed to the $[19, 23, 18]$ in the middle of the network and then decoded back to the initial size.

The first part, encoder, takes an image as the input and gradually decreases its resolutions, using strided convolutions, and simultaneously increases the number of channels. Each block in the encoder, except for the initial one, consists of two convolutions with kernel size $3 \times 3 \times 3$, preceded by instance normalization and ReLU activation layer. Downsampling blocks differ only in the sense that the first convolution has stride 2, instead of 1. Blocks have ResNet-like structure \citep{guerrero_white_2017} when the input is added to the output of the block.

Decoder, on the other side, steadily increases the spatial resolution of the image to return it to the initial size. Upsampling block does not have a residual connection, and it consists of one 3D convolution with kernel 3 preceded by instance normalization and ReLU activation and is followed by a trilinear upsampling with factor 2. Simple blocks are identical to the encoder part, except that they take as input not only the output of the previous layer but also an output of the encoder block with the same resolution. This feature of the U-Net model, known as skip-connection, allows the model to keep more details in the reconstruction.

\subsection{Deep Bayesian Models}  \label{sec:bayes}

In this part, we present deep Bayesian Inference and apply it to the U-Net model. Consider a U-Net model with $L$ convolutional layers. Denote by $w^{(i)}$, $i = 1,...,L$ kernels for the $i$th convolutional layer  and $w = (w^{(1)}, \ldots, w^{(L)})$ vector of all the model parameters. If kernel filters at a layer $i$ are of size $3\times 3\times 3$,  with $C_{inp}^{(i)}$ input channels and  $ C_{out}^{(i)}$ output channels, then the weight matrix has dimensions of  $ C_{inp}^{(i)} \times C_{out}^{(i)} \times 3\times 3\times 3$.

In the Bayesian approach, one combines prior distribution $p(w)$  on the parameters $w$ of the model with the information from observed training dataset $\mathcal{D} = \{(x_i, y_i)\}_{i=1}^N$ in the form of likelihood $p(\mathcal{D} |w)$ by posterior distribution $p(w|\mathcal{D} )$, computed with Bayes formula:

\begin{equation*}
p(w|\mathcal{D} ) = \frac{p(\mathcal{D} |w)p(w)}{p(\mathcal{D} )}.
\end{equation*}

For most cases, posterior distribution cannot be computed in closed form, since denominator of the above formula is not tractable. A common way to deal with this problem is to apply variational inference \citep{jordan1999introduction} when posterior is approximated by parametric distribution $q_{\theta}(w)$ which minimizes Kullback–Leibler divergence  between the true posterior $p(w|\mathcal{D} )$  and its variational approximation $q_{\theta}(w)$. More specifically, we are not interested in a point estimate of the model's weights $w$. Instead we are going to receive its distribution which is parametrized by $\theta$.

Moreover, we assume that both variational approximation $q_{\theta}(w)$ and prior distribution $p(w)$ are factorized over layers, input and output channels:

\begin{align*}
    q_{\theta}(w) &= \prod_{i = 1}^L \prod_{p = 1}^{C_{inp}^{(i)}} \prod_{k = 1}^{C_{out}^{(i)}} q_{\theta_{i p k}}(w^{(i)}_{p, k}),\\
    p(w) &= \prod_{i = 1}^L \prod_{p = 1}^{C_{inp}^{(i)}} \prod_{k = 1}^{C_{out}^{(i)}} p(w^{(i)}_{p, k}).
\end{align*}

Given all the assumptions above, the task burns down to the maximization of evidence lower bound (ELBO) \citep{hoffman2013stochastic} with respect to parameters of variational posterior distribution $\theta$:

\begin{equation}\label{eq:elbo}
\max_{\theta} \mathcal{L}(\theta) \approx \max_{\theta} \;\; \mathcal{L_{D^*}} - \sum_{i, p, k} 
 \underbrace{\int q_{\theta_{i p k}}(w^{(i)}_{p, k}) \log \left(\frac{q_{\theta_{i p k}}(w^{(i)}_{p, k})}{ p(w^{(i)}_{p, k})}\right)}_{\text{KL-divergence between }q_{\theta_{i p k}}(w^{(i)}_{p, k}) \text{ and } p(w^{(i)}_{p, k})}.
\end{equation}

Detailed derivation of the above expression is presented in Appendix A. The first part of the formula is a data term $\mathcal{L_{D^*}}$, also referred to as a reconstruction error. It is in charge of prediction quality, forcing the model to fit the data. Second term --- Kullback–Leibler divergence between variational distribution and prior over parameters of the model requires posterior distribution to be as close as possible to the prior, serving among other things as a regularization.

In the Bayesian framework, prior distribution is used to incorporate some knowledge or specific property, such as sparsity \citep{neklyudov2017structured} into parameters of the model. In the context of the current work, we consider prior distribution as a method for knowledge transfer. During our experiments with MRI semantic segmentation, we have noticed that kernels from different segmentation networks share a similar structure, when appropriately trained, in contrast to noisy kernels from models trained on small datasets. Therefore, prior distribution, which restricts kernels to be more structured, presumably should improve segmentation quality on modest training sets. We propose to apply Deep Weight Prior, discussed in the next part, to enforce precisely this property.

\subsection{Deep Weight Prior}  \label{sec:dwp}

Deep Weight Prior \citep{atanov_deep_2018} is an expressive prior distribution, which helps to incorporate information about the structure of previously learned convolutional filters during training of a new model. Prior is learned in the form of a generative model --- Variational Autoencoder \citep{kingma_vae}. It allows us to learn expressive distribution over the kernels, but we do not have direct access to its density and are only able to obtain samples.

Priors, whose probability density function (pdf) $p(w)$ is not accessible directly are called implicit in contrast to explicit priors, where pdf is available. To work with implicit priors we introduce some latent variables, assuming that conditional distribution with respect to them comes from some parametric family e.g., Gaussian distribution. We will use this method to work with Deep Weight Prior.

More precisely, we will consider implicit prior distribution in the form of Variational Autoencoder (VAE) \citep{kingma_vae} with encoder $r_{\psi^{(i)}}(x| w)$ and decoder $p_{\phi^{(i)}}(w|z)$, modeled by neural networks. Finally, given the prior over latent space $p(z)$, we arrive at the prior distribution for the kernels from the layer $i$:

\begin{equation*}
    p^{(i)}(w) = \int p_{\phi^{(i)}}(w|z) p(z)dz,
\end{equation*}

The main advantage of this prior is that it is non-restrictive, learnable from data and provides a fast sampling opportunity. Unfortunately, with implicit prior, it is not possible to compute Kullback–Leibler divergence from the ELBO objective (equation \ref{eq:elbo}). To this end, we follow the work of \cite{atanov_deep_2018} which replace KL-divergence by its upper bound.

\begin{equation*}
\text{KL}(q_{\theta}(w)|| p(w)) \leq \text{KL}^{\text{approx}},
\end{equation*}
    
\begin{equation*}
\text{KL}^{\text{approx}}_{i, p, k} = 
    -\mathbb{H}(q_{\theta_{i p k}}(w_{p,k}^{(i)})) + \mathbb{E}_{q_{\theta_{i p k}}(w_{p,k}^{(i)})}
    \left[
    \text{KL}( r_{\psi^{(i)}}(z| w^{(i)}_{p, k}) || p^{(i)}(w) ) - 
    \mathbb{E}_{r_{\psi^{(i)}}(z| w^{(i)}_{p, k})}
    \log p_{\phi^{(i)}}(w^{(i)}_{p, k}|z)) \right],
\end{equation*}

where $\mathbb{H}(\cdot)$ is an entropy of a corresponding distribution.

If $r_{\psi^{(i)}}(x| w)$, $p_{\phi^{(i)}}(w|z)$ and $q_{\theta}(w)$ are explicit distributions, we can use approximate lower bound (equation \ref{eq:elbo_approx}), for which we will be able to compute stochastic gradients with reparametrization trick to perform stochastic variational inference. We maximize approximate ELBO with respect to the parameters of the variational posterior distribution $\theta$ and DWP encoder parameters $\psi$. 

\begin{equation} \label{eq:elbo_approx}
  \max_{\theta, \psi} \mathcal{L}(\theta)^{\text{approx}} =\max_{\theta, \psi}\;\;  \mathcal{L_D} - \text{KL}^{\text{approx}}.
\end{equation}

\begin{figure}
\centering
\includegraphics[width=15cm]{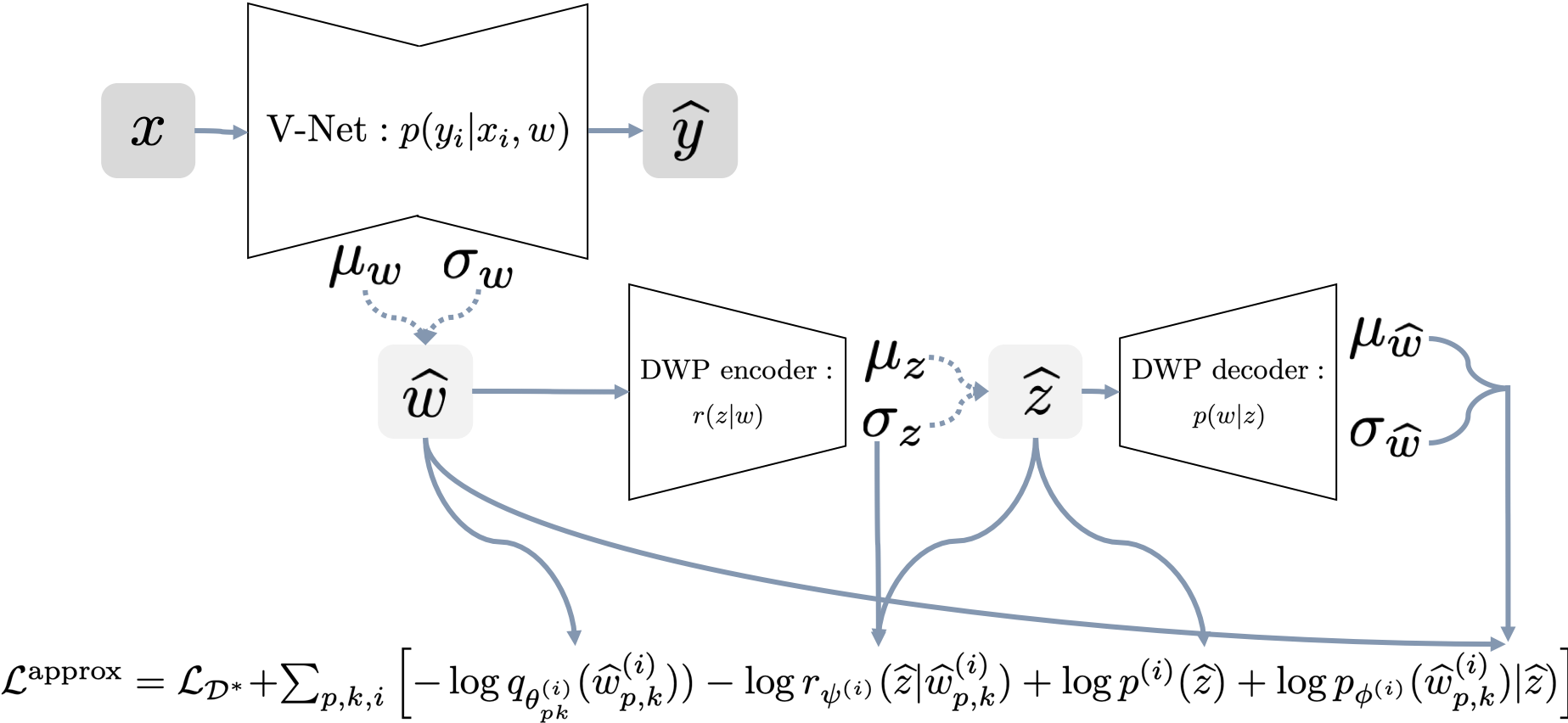}
\caption{Full scheme of the training procedure with Deep Weight Prior.}
\label{fig:dwp}
\end{figure}

The Algorithm \ref{alg:dwp} provides a pseudocode for the proposed algorithm. The algorithm requires as input the trained variational autoencoder on the reference dataset. We discuss particular details of training in section \ref{sec:exp}. Details on how different parts of the loss function are calculated, are presented in the Figure \ref{fig:dwp} for better understanding. We begin with sampling weights with reparametrization from variational distribution, which is fully factorized Gaussian $\widehat{w} \sim q_{\theta}(w)$. These samples are used to compute log-density of the variational posterior and parameters of the distribution $r_{\psi^{(i)}}(z|\widehat{w})$. Distribution $r_{\psi^{(i)}}(z|\widehat{w})$ is used to sample with reparametrization latent variable $\widehat{z} \sim r_{\psi^{(i)}}(z|\widehat{w})$ to further pass it to the decoder and obtain parameters of the distribution $\log p_{\phi} (\widehat{w}|\widehat{z})$. At this point, we have all the components of the objective to calculate stochastic gradient and update parameters $\theta$ of the U-Net and $\psi$ of the DWP encoder.

\begin{algorithm}
    \caption{Algorithm for training model with Deep Weight Prior.}
    \label{alg:dwp}
    \textbf{Input:} Dataset $\mathcal{D}: \,\{(x_i, y_i)\}_{i=1}^N$ \\
    \textbf{Input:} Variational approximation of the posterior distribution $q_{\theta}(w)$ \\
    \textbf{Input:} DWP with encoder $r_{\psi}(z|w)$ and decoder $p_{\phi} (w)$
    \begin{algorithmic}
        \While{not converged}
            \State Sample minibatch: $\mathcal{D^*}: \,\{(x_i, y_i)\}_{i=1}^M$
            \For{Layer $i \in \{1, \dots L\}$, input channel $p \in \{1, \dots C_{inp}\}$ and output channel $k \in \{1, \dots C_{out}\}$}
                \State Sample weights with reparametrization: $\widehat{w}_{pk}^{(i)} \sim q_{\theta_{i p k}}(w_{pk}^{(i)})$ 
                \State Sample latent variables with reparametrization: $\widehat{z}_{pk}^{(i)} \sim r_{\psi^{(i)}}(z|\widehat{w}_{pk}^{(i)}) $
            \EndFor
            \State Compute stochastic gradients of the objective: \\
    $ \mathcal{L}^{\text{approx}} = \mathcal{L_{M}} + \sum_{p, k, i} 
    \left[
    -\log q_{\theta_{i p k}}(\widehat{w}^{(i)}_{p, k}))
    - \log r_{\psi^{(i)}}(\widehat{z}| \widehat{w}^{(i)}_{p, k})
    + \log p(\widehat{z})
    + \log p_{\phi^{(i)}} (\widehat{w}^{(i)}_{p, k} | \widehat{z})
    \right]$ \\
            \State Update parameters $\theta = \theta + \alpha \bigtriangledown_{\theta} \mathcal{L}^{\text{approx}}$ and $\psi =\psi + \beta  \bigtriangledown_{\psi} \mathcal{L}^{\text{approx}}$
        \EndWhile
    \end{algorithmic}
\textbf{Output:} $q_{\theta}(w)$ --- posterior distribution of the model parameters
\end{algorithm}

\subsection{Datasets} \label{sec:dataset}
To emphasize the ability of the proposed approach to generalizing, two public available datasets were chosen with different diseases on the challenging task of the brain segmentation. 

\begin{figure}
\centering
\begin{subfigure}{.45\textwidth}
  \centering
  \includegraphics[width=0.9\textwidth]{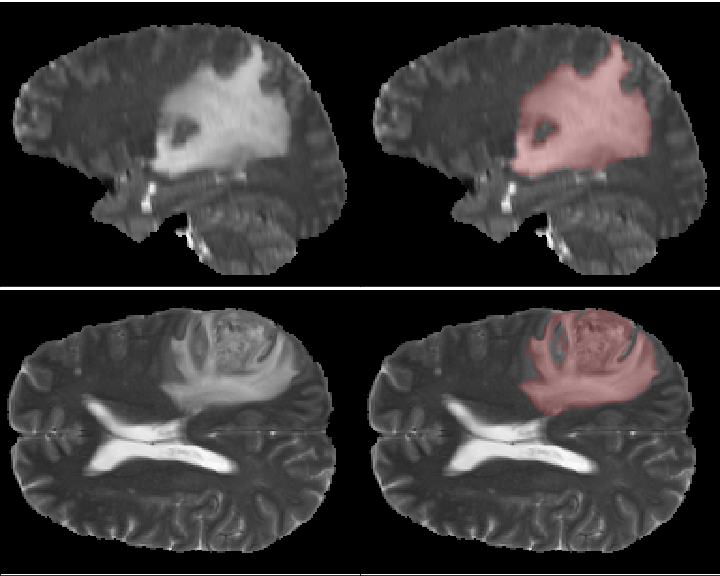}
  \caption{BRATS18 dataset.}
  \label{fig:brats_ex}
\end{subfigure}%
\begin{subfigure}{.45\textwidth}
  \centering
  \includegraphics[width=0.9\textwidth]{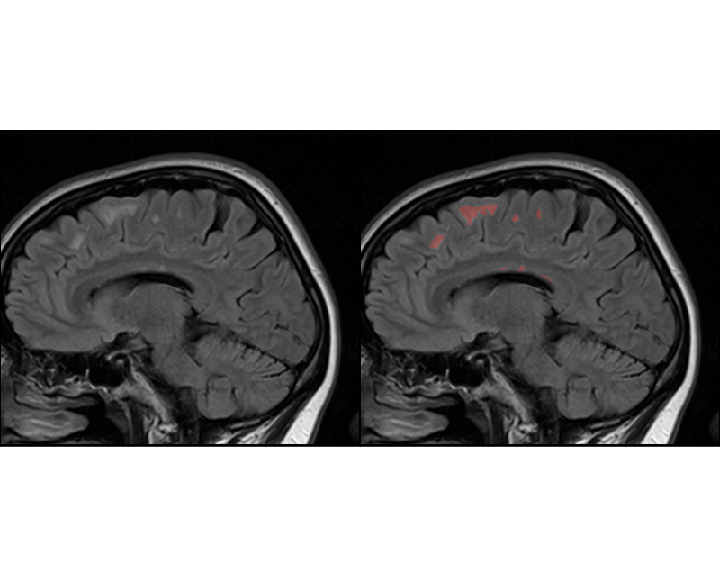}
    \caption{MS dataset.}
    \label{fig:ms_ex}
    \end{subfigure}
\caption{Example of MRI slices and ground truth segmentation.}
\label{fig:test}
\end{figure}

First one comes from the annual competition on brain tumor segmentation, BRATS18 \citep{menze_multimodal_2015, bakas2017advancing}. It contains pre-operative MRI scans of 275 patients with glioblastoma (GBM/HGG) and lower grade glioma (LGG). Each volume has resolution 240 $\times$ 240 $\times$ 155 pixels, acquired with different protocols and scanners in 19 institutions. All the images were co-registered, reshaped to the same resolution and skull-stripped. Ground truth labels were manually created by expert neuroradiologists for all the scans. The analysis was performed on T2-weighted volumes. Figure \ref{fig:brats_ex} shows an example from this dataset. The second dataset is Multiple Sclerosis Human Brain MR Imaging Dataset (MS) \citep{MsDataset}, which is available on the Skoltech CoBrain Analytics platform. This dataset contains 170 manually labelled MRI FLAIR sequences of subjects with multiple sclerosis. All the images were acquired on 1.5T Siemens Magnetom Avanto scanner with slice thickness = 5 mm, slice spacing = 1.5 mm and have resolution 448 $\times$ 512 $\times$ 22. Figure \ref{fig:ms_ex} depicts one sample from this dataset.

\subsection{Evaluation methods} \label{sec:eval}
Two conventional metrics for semantic segmentation \citep{clerigues_sunet:_2018, kao_brain_2018, myronenko_3d_2018, deniz_segmentation_2018} are used to evaluate the model performance --- Dice Similarity Coefficient (DSC), also known as $F_1$-score, and Intersection over union (IoU):

\begin{align*}
\text{DSC} &= \frac{2TP}{2TP + FN + FP} = \frac{2\text{IoU}}{1 + \text{IoU}},\\
\text{IoU} &= \frac{TP}{TP + FN + FP}.
\end{align*}

The number of true positive (TR), false negative (FN) and false positive (FP) pixels is calculated for each image and averaged over the whole test set. We compare model predictions with the segmentations, which have been manually created by professional radiologists \citep{menze_multimodal_2015} and are considered as ground truth.

\subsection{Loss}

To train U-Net in the non-Bayesian setting, we use a combination of binary cross-entropy and Dice losses. We apply this loss when training all models without Deep Weight Prior: for pre-training on the source dataset, fine-tuning on target dataset and training models with random initialization. 

The first component of the loss, binary cross-entropy, is a common loss function for classification problem \citep{goodfellow2016deep}. In semantic segmentation setting we classify each pixel of the input image, resulting in the following loss function:

\begin{equation*}
    \mathcal{L}_{CE} = - \sum_{i, j} y_{i,j} \log \widehat{y}_{i,j} +  (1-y_{i,j}) \log (1-\widehat{y}_{i,j}),
\end{equation*}
where $\widehat{y}_{i,j}$ is a predicted probability for pixel $j$ from image $i$ to be from the class of interest. Problem with cross-entropy is that it does not account for class imbalance, which usually takes place in semantic segmentation tasks, since background is the most prevalent class. Dice loss, in contrast, is known to be robust to this problem. It is based on Dice Similarity Coefficient and defined as:

\begin{equation*}
    \mathcal{L}_{DICE} = \sum_{i = 1}^N \frac{FN_i + FP_i}{2TP_i + FN_i + FP_i} =  \sum_{i = 1}^N (1 - DSC^{(i)}).
\end{equation*}

The weight of each component in the final combination was chosen experimentally. Since cross-entropy loss resulted in model learning to predict background for all the pixels in most cases, we arrived to the setting where it has a low weight of 0.01:

\begin{equation*}
    \mathcal{L} =0.99\cdot \mathcal{L}_{DICE} + 0.01\cdot \mathcal{L}_{CE}.
\end{equation*}

\subsection{Experimental setup} \label{sec:exp}

The aim of the experiments is to compare the proposed method (Unet-DWP) with the conventional transfer learning approach: training the model on the small target dataset with pretrained on the source dataset (UNet-PR) or freezing layers in the middle of the network (UNet-PRf) while fine-tuning only the first and the last block of the model to reduce overfitting on a small dataset. As a baseline, we also consider random initialization (UNet-RI), where the model is trained only on the small target dataset. We use initialization introduced in \cite{he2015delving}, also known as He initialization for UNet-RI. The training procedure summarized in the Algorithm \ref{alg:train_pre-train} for pre-training approaches and in the Algorithm \ref{alg:train_dwp} for proposed training with deep prior (UNet-DWP). To compare the proposed methods, we use MS dataset as a source and small subsets of BRATS18 dataset as targets. Both dataset consider the MRI scans of the brain, however with different diseases. The purpose of this setup to show the ability of the method to generalize between diseases. Models performance was compared on the whole tumour segmentation on subsets of BRATS18 volumes, containing 5, 10, 15 or 20 randomly selected images with the fixed test sample size of 50 images. The proposed method is mostly relevant for datasets of small sizes since they do not contain enough samples to train proper network and prior knowledge from a larger dataset should improve the quality.

\begin{algorithm}
    \caption{Procedure for UNet-PR training on $m$ images.}
    \label{alg:train_pre-train}
    \textbf{Input:} Dataset to train prior on  $\mathcal{D}_{prior}: \,\{(x_i, y_i)\}_{i=1}^N$ 
    
    \textbf{Input:} Target dataset $\mathcal{D}_{target}: \,\{(x_i, y_i)\}_{i=1}^n$ 
    \begin{algorithmic}
    \State Train one 3D U-Net model on $\mathcal{D}_{prior}$ and remember the weights
    \For{Iteration $\in [1, 2, 3]$}
        \State Split $\mathcal{D}_{target}$ on train and test: $\mathcal{D}_{target}^{Train}$, $\mathcal{D}_{target}^{Test}$
        \State Select $m$ images from $\mathcal{D}_{target}^{Train}$
        \State Initialize model with the weights trained on $\mathcal{D}_{prior}$
        \State Train 3D U-Net on selected images
        \State Evaluate model on $\mathcal{D}_{target}^{Test}$
    \EndFor
    \end{algorithmic}
\textbf{Output:} Trained 3D U-Net model
\end{algorithm}

\subsubsection{U-Net Training Details}
All the models on the target dataset were trained on the whole volumes with batch size 2 and without any data augmentation.  Table \ref{table:hyperparams} summarize hyperparameters details used during training. For training we use Adam optimizer with initial learning rate $10^{-3}$. Learning rate is decreased by the factor of 10, when loss on the validation set is not decreasing by more than $10^{-4}$ during 10 epochs. We stop training the model as soon as learning rate reaches the value $10^{-6}$. Three different train-test splits of BRATS18 were used for validation in order to verify the robustness of the result. All the experiments were performed on the NVIDIA Tesla V100-SXM2 GPUs.

\begin{table}
\begin{center}
\begin{tabular}{@{}ll@{}}
\toprule
Parameter   & Value            \\ \midrule
Batch-size             & 2 \\
Optimizer   & Adam \\
Initial Learning rate & $10^{-3}$ \\
LR scheduler & Reduce learning rate when a loss has stopped improving \\
LR scheduler patience &  10 \\
LR scheduler factor &  0.1\\ 
Max epochs & 500 \\
Early stopping criterion & LR == $10^{-6}$\\
Test size & 50 \\
Train sizes & [5, 10, 15, 20] \\
\bottomrule
\end{tabular}
\caption{UNet hyperparameters details.}
\label{table:hyperparams}
\end{center}
\end{table}

Kernels for further DWP training were collected from UNet network, trained on the while volumes of the source dataset. Batch size, optimizer and LR scheduler are presented in Table~\ref{table:hyperparams}. We have applied this setting to train 10 models until convergence, which took on average 100 epochs for one model. To obtain more filters, we have applied cyclical learning rate \citep{smith2017cyclical} to obtain 10 more networks. That is we increase learning rate back to $10^{-3}$ for a converged model and continue training it with the same LR scheduler to converge to a new minimum. As a result, we end up with 20 trained networks with average Dice Score of 0.61 on validation set. As can be seen from the Figure~\ref{fig:dwp_ker}, obtained filters have clear structure, which indicates they potential usefulness.

\subsubsection{DWP training details} 
To train the DWP prior we should specify the number and architecture of the variational auto-encoders and collect the training set of filters. We train variational autoencoder with latent vector dimention 6. We've used Adam optimizer, batch size of 20 images. All the hyperparameters are presented in the Table~\ref{table:hyperparams_dwp}. Appendix B contains architectures, which were used. We assume that filters from the layers, which take as input images of the same resolution come from the same distribution. As a result, seven Variational Autoencoders were trained and served further as implicit prior distributions for the kernels of the corresponding layers. To obtain the training set of filters U-Net models were trained on the whole MS dataset with random initialization. Afterwards, kernels were collected from trained models to train prior in the form of the Variational Autoencoder.

\begin{table}
\begin{center}
\begin{tabular}{@{}ll@{}}
\toprule
Parameter   & Value            \\ \midrule
Batch-size             & 20 \\
Optimizer   & Adam \\
Initial Learning rate & $10^{-3}$ \\
LR scheduler & Reduce learning rate when a loss has stopped improving \\
LR scheduler patience &  15 \\
LR scheduler factor &  0.1\\ 
Max epochs & 500 \\
Early stopping criterion & LR == $10^{-6}$\\
Latent dimension & 6 \\
\bottomrule
\end{tabular}
\caption{DWP hyperparameters details.}
\label{table:hyperparams_dwp}
\end{center}
\end{table}

\begin{algorithm}\label{algo_unet_dwp}
    \caption{Procedure of UNet-DWP training on $m$ images.}
    \label{alg:train_dwp}
    \textbf{Input:} Dataset to train prior on  $\mathcal{D}_{prior}: \,\{(x_i, y_i)\}_{i=1}^N$ 
    
    \textbf{Input:} Target dataset $\mathcal{D}_{target}: \,\{(x_i, y_i)\}_{i=1}^n$ 
    \begin{algorithmic}
    \State Train 3D U-Net models with different initializations on $\mathcal{D}_{prior}$
    \State Collect kernels and split them into seven parts (depending on the input size of the layer)
    \State Train 7 VAE, to use them as implicit prior
    \For{Iteration $\in [1, 2, 3]$}
        \State Split $\mathcal{D}_{target}$ on train and test: $\mathcal{D}_{target}^{Train}$, $\mathcal{D}_{target}^{Test}$
        \State Select $m$ images from $\mathcal{D}_{target}^{Train}$
        \State Train 3D U-Net on selected images with Deep Weight Prior
        \State Evaluate model on $\mathcal{D}_{target}^{Test}$
    \EndFor
    \end{algorithmic}
\textbf{Output:} Trained 3D U-Net model
\end{algorithm}

\section{Results} \label{sec:result}
Each model (UNet-RI, UNet-DWP, UNet-PR and UNet-PRf) was estimated at three different random train/test splits. For a fixed test sample of 50 images 5, 10, 15 and 20 images were selected for training, and on each sample, three models were estimated. Tables \ref{table:results_dsc} and  \ref{table:results_iou}  summarize the obtained results. UNet-RI stands for the model trained with the random initialization, UNet-PR and UNet-PRf are transfer learning approaches (in the second case, weights of the middle layers were frozen), where U-Net was pre-trained on MS dataset and, finally, UNet-DWP is a model trained with Deep Weight Prior. We calculate mean DSC and IoU metrics for different train-test splits and its standard deviation, which is given in the brackets.

 We can see that models trained with DWP noticeably outperformed both randomly initialized and pre-trained U-Net for all the training sizes. We observe higher variability in prediction accuracy for the problems with smaller sample sizes, which shrinks as training dataset grows, and the superiority of UNet-WDP becomes clearer. It is also worth mentioning that the pre-trained mode where part of the weights were frozen fails. We believe that this means that information from other diseases is not relevant for the new task by default, and without fine-tuning of the whole network, we are not able to achieve consistent results.

\begin{figure}[h]
\centering
\includegraphics[width=13cm]{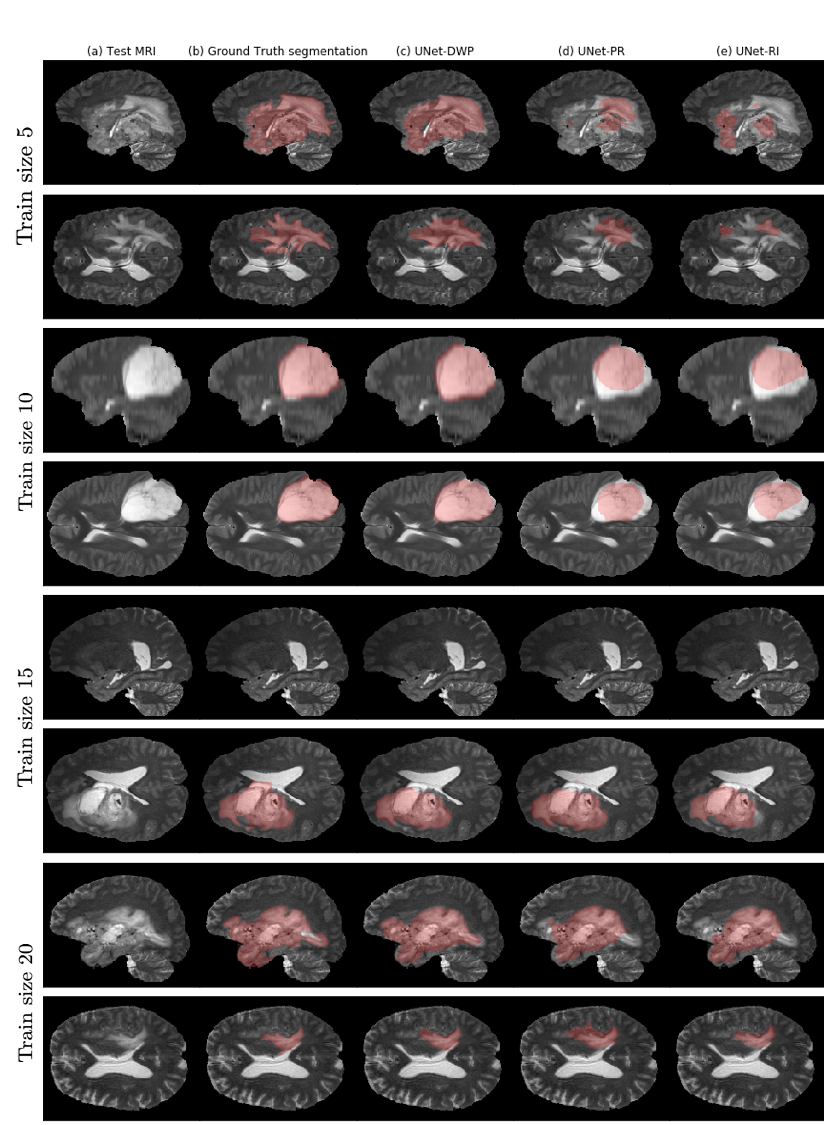}
\caption{Examples of models' predictions on test samples, compared to ground truth segmentation}
\label{fig:preditions}
\end{figure}

 Figure \ref{fig:preditions} contains example predictions of different models (panels c-e) along with ground truth segmentations (panel b). Each row corresponds to different training sample size. For example, for the model trained on 10 images, there is a notable difference in tumor coverage for UNet-DWP and UNet-PR models, which results in DSC of 0.92 for the first model and 0.74 for the second. On other images we may also note, that model with DWP manages to cover more relevant areas.

\begin{table}
\begin{center}
\begin{tabular}{@{}lllllllll@{}}
\toprule
Train size  & UNet-DWP (ours)          & UNet-PR   & UNet-PRf      & UNet-RI     \\ \midrule
5          &\textbf{0.64} (0.05) & 0.61 (0.02)& 0.58 (0.03) & 0.62 (0.02)\\
10         & \textbf{0.71} (0.04) & 0.64 (0.01)& 0.60 (0.03) & 0.66 (0.01)  \\
15         & \textbf{0.71} (0.02) & 0.67 (0.02)&0.63 (0.02)  & 0.70 (0.02) \\
20         &\textbf{0.74} (0.01) & 0.69 (0.01)& 0.65 (0.02)  & 0.70 (0.01) \\ \bottomrule
\end{tabular}
\caption{Mean Dice Similarity Score for the experiments with small available target dataset.}
\label{table:results_dsc}
\end{center}
\end{table}

\begin{table}
\begin{center}
\begin{tabular}{@{}lllllllll@{}}
\toprule
Train size  & UNet-DWP (ours)           & UNet-PR   & UNet-PRf      & UNet-RI     \\ \midrule
5         &\textbf{0.52} (0.05) & 0.49 (0.02)& 0.45 (0.03) & 0.50 (0.02) \\
10        & \textbf{0.58} (0.05) & 0.52 (0.01)& 0.47 (0.03) & 0.53 (0.01) \\
15         & \textbf{0.60} (0.02) & 0.56 (0.02)& 0.50 (0.02) & 0.58 (0.02) \\
20        & \textbf{0.63}(0.01) & 0.58 (0.01)& 0.53 (0.02) & 0.60 (0.01) \\ \bottomrule
\end{tabular}
\caption{Intersection over Union metrics for the experiments with small available target dataset.}
\label{table:results_iou}
\end{center}
\end{table}

It is worth mentioning, that transfer learning model on average performs even worse than the model without any prior knowledge about the data. This result is quite surprising, but it can be explained by strong disease specificity of the data. Even from the examples in Figures \ref{fig:brats_ex}, \ref{fig:ms_ex} it can be seen, that datasets differ not only in the shapes of the target segmentation (plaques of multiple sclerosis are much smaller and difficult to notice that brain tumor) but also in resolution, contrast and preprocessing method, as a result, after corresponding initialization, fine-tuning may converge to a worse solution.

\begin{figure}
\centering
\includegraphics[width=\textwidth]{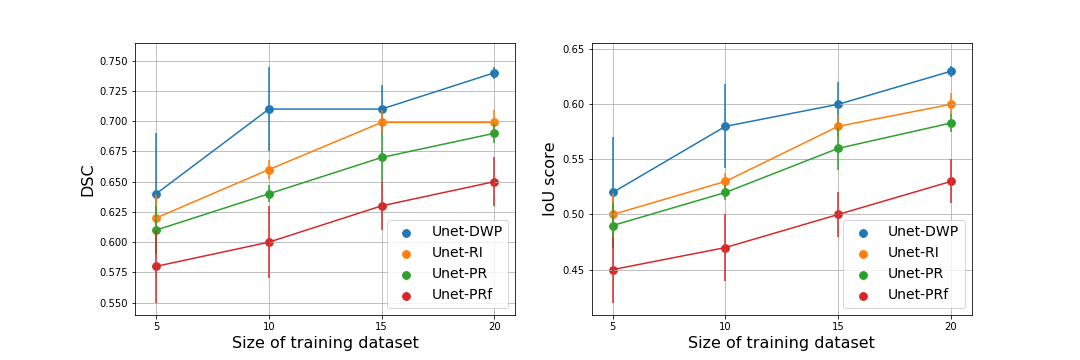}
\caption{Segmentation accuracy on BRATS18 dataset for various train sample size, calculated for three different splits.}
\label{fig:res}
\end{figure}

\begin{figure}
\centering
\begin{subfigure}{.45\textwidth}
  \centering
  \includegraphics[width=0.6\textwidth]{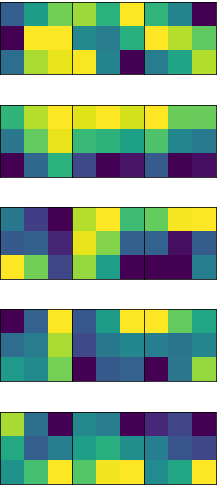}
  \caption{Kernels from U-Net, trained on MS dataset.}
  \label{fig:dwp_ker1}
\end{subfigure}%
\begin{subfigure}{.45\textwidth}
  \centering
  \includegraphics[width=0.6\textwidth]{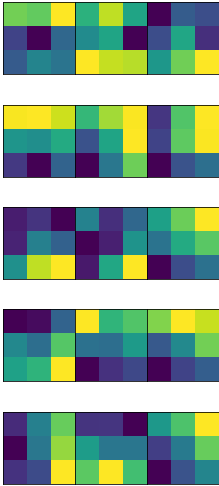}
    \caption{Samples from trained Deep Weight Prior.}
    \label{fig:dwp_ker2}
    \end{subfigure}
\caption{Examples of trained kernels.}
\label{fig:dwp_ker}
\end{figure}

Figure \ref{fig:dwp_ker} illustrates prior over the weights, that was used for the given experiment. Panel a) contains kernels of the U-Net, trained on the MS dataset. Since the dataset is big enough, they are not noisy and have clear structure, as it was expected. Panel b) depicts samples from Variational Autoencoder, which was later used as an implicit prior distribution. Even though samples from the Deep Weight Prior on the right are not identical to the real kernels on the left, they still have similar structure and we can assume that the VAE managed to grasp a proper distribution.

\section{Discussion} \label{sec:discussion}
The proposed method can be used for knowledge transfer between medical imaging data from different domains, resulting in performance improvement over conventional transfer learning. This method is based on the deep Bayesian approach and exploits implicit prior distribution over convolutional filters. 

Our approach is not limited to a specific task and can be applied to such problems as classification, detections or any other, where convolutional neural networks are used. But we believe that it is the most relevant for the semantic segmentation problem. There are plenty of challenges in this area. One of the most significant is that manual segmentation of MRI volumes, which is needed to train any supervised model, is very expensive. The reason is that it requires the work of several professional radiologists and each image should be labelled several times by different people to reduce ambiguity. For instance, it takes around 60 minutes for a radiologist to label one scan of the patient with a brain tumor, resulting in 4 hours of work per observation \citep{menze_multimodal_2015}.
Moreover, institutions are often hesitant to share data with external collaborators because of patient privacy, as well as ethical and legal considerations. As a result, there are very few publicly available datasets, and they are often tiny, up to 5 or 10 images. Besides, data is highly disease-specific, making conventional transfer learning technique inefficient for cases, when source dataset, used for initial model training, has a different domain (another illness, MRI modality and preprocessing method), which is confirmed in our experiments.

The most popular model for semantic segmentation is volumetric U-Net \citep{deniz_segmentation_2018, livne_u-net_2019, guerrero_white_2017}. The idea behind this architecture is quite simple; it is based on conventional U-Net model for semantic segmentation. The main advantage of this models is that it proceeds the whole 3D image, using 3D convolutions, instead of working with 2-dimensional slices separately. It is also quite common to use different heuristic regularization techniques: VAE regularization, \citep{myronenko_3d_2018}, anatomically constrained U-Net \citep{oktay_anatomically_2018, degel_domain_2018}.

In this work, we compare three methods for semantic segmentation of a brain tumour on small datasets of size from 5 to 20. In the first approach, we train 3D U-Net from scratch, using the information only from a given small dataset, in the second approach we firstly train a model on a large dataset with multiple sclerosis segmentation and then use trained kernels to initialize model for brain tumour segmentation. Finally, we propose a new approach to transfer information using Deep Weight Prior --- implicit prior distribution, also learned on a large dataset with multiple sclerosis and applied to train tumour segmentation network. We have shown that the proposed approach outperforms both simple and fine-tuned models. Presumably, transfer learning approach fails in this case because prior was trained on the samples with different illness and information about it is not a proper initialization for a new task. At the same time, Deep Weight Prior ensures that kernels come from similar distribution, bringing up structure into convolutional filters. Even though all the experiments were performed on a simple U-Net model, it can be applied to any other architecture with a more complicated structure.

\subsection{Extra experiments} 
During our experiments, we aimed at using datasets containing the same organs to make sure that the source data has enough relevant information to transfer to the target one. But of course, the proposed method is not limited to the specific part of the human body and can be applied to other organs as well. To test this hypothesis, we performed additional experiments on the dataset, containing CT scans of the liver from the medical decathlon segmentation challenge  \citep{simpson2019large}. As a source \texttt{Task08\_HepaticVessel} dataset was used, containing 443 CT scans of patients with liver tumours. As a target dataset subsets of \texttt{Task03\_Liver} and  \texttt{Task09\_Spleen} datasets were used. The first target dataset is closer to the source one since it contains 201 CT images of patients with a liver tumour. In the second dataset the aim is to segment spleen on the CT scans for 41 patients, which makes transfer learning from the source dataset a more challenging task.

The purpose of this additional experiment was to provide evidence that the proposed method can be successfully applied not only to the brain but also to other organs. We did not tune the architecture to reach state-of-the-art performance for the specific dataset. Instead, we applied the same architecture, experimental setups as in the main part of the paper. The only difference was that due to the large image size, we had to use patches instead of the whole volumes ([40, 400, 400] for the source, [192, 192, 192] for the first target dataset and [24, 480, 480] for the second target dataset).

Preliminary results that we have obtained are quite promising. They are presented in table \ref{table:results_liver_dsc} and show that for both target datasets UNet with Deep Weight Prior performs better than the competitors in most cases. We believe that this part could be further improved, by tuning the architecture and adding more datasets for comparison.

\begin{table}[h]
\begin{center}
\begin{tabular}{@{}llll|lll@{}}
\toprule
          & \multicolumn{3}{c}{\texttt{Task03\_Liver}}         & \multicolumn{3}{c}{\texttt{Task09\_Spleen}} \\
Train size & UNet-DWP (ours)       & UNet-RI        & UNet-PR & UNet-DWP (ours)       & UNet-RI        & UNet-PR \\\midrule
5          & 0.275          & \textbf{0.284} & 0.209   & \textbf{0.467} & 0.391          & 0.105   \\
10         & \textbf{0.328} & 0.293          & 0.052   & \textbf{0.625} & 0.584          & 0.239   \\
15         & \textbf{0.389} & 0.306          & 0.243   & 0.556          & \textbf{0.579} & 0.302   \\
20         & \textbf{0.353} & 0.336          & 0.156   & \textbf{0.649} & 0.566          & 0.459 \\ \bottomrule
\end{tabular}
\caption{Mean Dice Similarity Score for the subsets of \texttt{Task03\_Liver} and \texttt{Task09\_Spleen} datasets.}
\label{table:results_liver_dsc}
\end{center}
\end{table}

The reasonable question arises is the necessity for transferring when a relatively large dataset is available. Hence, we consider the additional experiment of transfer learning from the MS dataset as the source to the BRATS18 as the target, while 100 of samples are available from the target dataset. 

\begin{table}[h]
\begin{center}
\begin{tabular}{@{}lllllllll@{}}
\toprule
Train size  & UNet-DWP (ours)          & UNet-PR   & UNet-PRf      & UNet-RI     \\ \midrule
100         &0.76 (0.01) & 0.79 (0.01)& 0.77 (0.01)  & 0.77 (0.01) \\ \bottomrule
\end{tabular}
\caption{Mean Dice Similarity Score for the experiments with large available target dataset (MS-BRATS18).}
\label{table:results_dsc_100}
\end{center}
\end{table}

Taking into account standard deviation, results are quite close to each other. As it was expected, our method converges to the UNet-RI result as a number of training samples increases, since in this case knowledge transfer becomes less useful because there is enough information in the target dataset to train a proper network \citep{lu2017bernstein}.

Further research on the topic may include experiments with knowledge transfer from other problem settings, e.g., from classification to segmentation and vice versa. The first setting is of higher interest, since there are usually more observations in classification datasets and there are more of them available for different diseases, making it a more accessible source of prior knowledge.

\section*{Funding}

This work was supported by the Ministry of Education and Science of the Russian Federation (Grant no. 14.756.31.0001)

The authors acknowledge the usage of the Skoltech CDISE HPC cluster ``Zhores'' for obtaining the results presented in this paper.

\section*{Acknowledgements}
Authors would like to thank Anh-Huy Phan for fruitful discussions and valuable advice during our work on this paper.



\bibliographystyle{plainnat}  
\bibliography{main}  

\begin{thebibliography}{52}
\providecommand{\natexlab}[1]{#1}
\providecommand{\url}[1]{\texttt{#1}}
\expandafter\ifx\csname urlstyle\endcsname\relax
  \providecommand{\doi}[1]{doi: #1}\else
  \providecommand{\doi}{doi: \begingroup \urlstyle{rm}\Url}\fi

\bibitem[Atanov et~al.(2018)Atanov, Ashukha, Struminsky, Vetrov, and
  Welling]{atanov_deep_2018}
Andrei Atanov, Arsenii Ashukha, Kirill Struminsky, Dmitry Vetrov, and Max
  Welling.
\newblock The {Deep} {Weight} {Prior}.
\newblock \emph{arXiv preprint arXiv:1810.06943}, 2018.
\newblock URL \url{http://arxiv.org/abs/1810.06943}.

\bibitem[Bakas et~al.(2017)Bakas, Akbari, Sotiras, Bilello, Rozycki, Kirby,
  Freymann, Farahani, and Davatzikos]{bakas2017advancing}
Spyridon Bakas, Hamed Akbari, Aristeidis Sotiras, Michel Bilello, Martin
  Rozycki, Justin~S Kirby, John~B Freymann, Keyvan Farahani, and Christos
  Davatzikos.
\newblock Advancing the cancer genome atlas glioma mri collections with expert
  segmentation labels and radiomic features.
\newblock \emph{Scientific data}, 4:\penalty0 170117, 2017.

\bibitem[Christodoulidis et~al.(2017)Christodoulidis, Anthimopoulos, Ebner,
  Christe, and Mougiakakou]{christodoulidis2017multisource}
Stergios Christodoulidis, Marios Anthimopoulos, Lukas Ebner, Andreas Christe,
  and Stavroula Mougiakakou.
\newblock Multisource transfer learning with convolutional neural networks for
  lung pattern analysis.
\newblock \emph{IEEE journal of biomedical and health informatics}, 21\penalty0
  (1):\penalty0 76--84, 2017.

\bibitem[{\c{C}}i{\c{c}}ek et~al.(2016){\c{C}}i{\c{c}}ek, Abdulkadir, Lienkamp,
  Brox, and Ronneberger]{cciccek20163d}
{\"O}zg{\"u}n {\c{C}}i{\c{c}}ek, Ahmed Abdulkadir, Soeren~S Lienkamp, Thomas
  Brox, and Olaf Ronneberger.
\newblock 3d {U-Net}: learning dense volumetric segmentation from sparse
  annotation.
\newblock In \emph{International Conference on Medical Image Computing and
  Computer-Assisted Intervention}, pages 424--432. Springer, 2016.

\bibitem[Cl{\`e}rigues et~al.(2018)Cl{\`e}rigues, Valverde, Bernal, Freixenet,
  Oliver, and Llad{\'o}]{clerigues_sunet:_2018}
Albert Cl{\`e}rigues, Sergi Valverde, Jose Bernal, Jordi Freixenet, Arnau
  Oliver, and Xavier Llad{\'o}.
\newblock {SUNet}: a deep learning architecture for acute stroke lesion
  segmentation and outcome prediction in multimodal mri.
\newblock \emph{arXiv preprint arXiv:1810.13304}, 2018.

\bibitem[CoBrain analytics()]{MsDataset}
CoBrain analytics.
\newblock Multiple sclerosis human brain {MR} imaging dataset.
\newblock
  \url{https://app.cobrain.io/datasets/2c683256-6dcd-47bc-9399-34e166c3fc71},
  2018.

\bibitem[Davatzikos et~al.(2008)Davatzikos, Resnick, Wu, Parmpi, and
  Clark]{davatzikos2008individual}
Christos Davatzikos, Susan~M Resnick, X~Wu, P~Parmpi, and Christopher~M Clark.
\newblock Individual patient diagnosis of {AD} and {FTD} via high-dimensional
  pattern classification of {MRI}.
\newblock \emph{Neuroimage}, 41\penalty0 (4):\penalty0 1220--1227, 2008.

\bibitem[Davuluri et~al.(2012)Davuluri, Wu, Tang, Cockrell, Ward, Najarian, and
  Hargraves]{davuluri_hemorrhage_2012}
Pavani Davuluri, Jie Wu, Yang Tang, Charles~H. Cockrell, Kevin~R. Ward, Kayvan
  Najarian, and Rosalyn~H. Hargraves.
\newblock Hemorrhage detection and segmentation in traumatic pelvic injuries.
\newblock \emph{Computational and Mathematical Methods in Medicine}, 2012.
\newblock \doi{10.1155/2012/898430}.
\newblock URL \url{https://www.hindawi.com/journals/cmmm/2012/898430/}.

\bibitem[Degel et~al.(2018)Degel, Navab, and Albarqouni]{degel_domain_2018}
Markus~A Degel, Nassir Navab, and Shadi Albarqouni.
\newblock Domain and geometry agnostic {CNNs} for left atrium segmentation in
  3d ultrasound.
\newblock \emph{arXiv preprint arXiv:1805.00357}, 2018.

\bibitem[Deniz et~al.(2018)Deniz, Xiang, Hallyburton, Welbeck, Babb, Honig,
  Cho, and Chang]{deniz_segmentation_2018}
Cem~M. Deniz, Siyuan Xiang, R.~Spencer Hallyburton, Arakua Welbeck, James~S.
  Babb, Stephen Honig, Kyunghyun Cho, and Gregory Chang.
\newblock Segmentation of the {Proximal} {Femur} from {MR} {Images} using
  {Deep} {Convolutional} {Neural} {Networks}.
\newblock \emph{Scientific Reports}, 8\penalty0 (1):\penalty0 16485, 2018.
\newblock ISSN 2045-2322.
\newblock \doi{10.1038/s41598-018-34817-6}.
\newblock URL \url{https://www.nature.com/articles/s41598-018-34817-6}.

\bibitem[Elsayed et~al.(2018)Elsayed, Goodfellow, and
  Sohl-Dickstein]{elsayed2018adversarial}
Gamaleldin~F Elsayed, Ian Goodfellow, and Jascha Sohl-Dickstein.
\newblock Adversarial reprogramming of neural networks.
\newblock \emph{arXiv preprint arXiv:1806.11146}, 2018.

\bibitem[Ghafoorian et~al.(2017)Ghafoorian, Mehrtash, Kapur, Karssemeijer,
  Marchiori, Pesteie, Guttmann, de~Leeuw, Tempany, van Ginneken,
  et~al.]{ghafoorian2017transfer}
Mohsen Ghafoorian, Alireza Mehrtash, Tina Kapur, Nico Karssemeijer, Elena
  Marchiori, Mehran Pesteie, Charles~RG Guttmann, Frank-Erik de~Leeuw, Clare~M
  Tempany, Bram van Ginneken, et~al.
\newblock Transfer learning for domain adaptation in mri: Application in brain
  lesion segmentation.
\newblock In \emph{International Conference on Medical Image Computing and
  Computer-Assisted Intervention}, pages 516--524. Springer, 2017.

\bibitem[Gong et~al.(2007)Gong, Liu, Tan, Farzad, Lee, Pang, Tian, Tang, and
  Zhang]{gong2007classification}
Tianxia Gong, Ruizhe Liu, Chew~Lim Tan, Neda Farzad, Cheng~Kiang Lee,
  Boon~Chuan Pang, Qi~Tian, Suisheng Tang, and Zhuo Zhang.
\newblock Classification of {CT} brain images of head trauma.
\newblock In \emph{IAPR International Workshop on Pattern Recognition in
  Bioinformatics}, pages 401--408. Springer, 2007.

\bibitem[Goodfellow et~al.(2016)Goodfellow, Bengio, and
  Courville]{goodfellow2016deep}
Ian Goodfellow, Yoshua Bengio, and Aaron Courville.
\newblock \emph{Deep learning}.
\newblock MIT press, 2016.

\bibitem[Guerrero et~al.(2018)Guerrero, Qin, Oktay, Bowles, Chen, Joules, Wolz,
  Vald{\'e}s-Hern{\'a}ndez, Dickie, Wardlaw, et~al.]{guerrero_white_2017}
R~Guerrero, C~Qin, O~Oktay, C~Bowles, L~Chen, R~Joules, R~Wolz,
  MC~Vald{\'e}s-Hern{\'a}ndez, DA~Dickie, J~Wardlaw, et~al.
\newblock White matter hyperintensity and stroke lesion segmentation and
  differentiation using convolutional neural networks.
\newblock \emph{NeuroImage: Clinical}, 17:\penalty0 918--934, 2018.

\bibitem[Hammers et~al.(2007)Hammers, Heckemann, Koepp, Duncan, Hajnal,
  Rueckert, and Aljabar]{hammers2007automatic}
Alexander Hammers, Rolf Heckemann, Matthias~J Koepp, John~S Duncan, Jo~V
  Hajnal, Daniel Rueckert, and Paul Aljabar.
\newblock Automatic detection and quantification of hippocampal atrophy on mri
  in temporal lobe epilepsy: a proof-of-principle study.
\newblock \emph{Neuroimage}, 36\penalty0 (1):\penalty0 38--47, 2007.

\bibitem[Han et~al.(2018)Han, Yoo, Kim, Shin, Sung, and Ye]{han2018deep}
Yoseob Han, Jaejun Yoo, Hak~Hee Kim, Hee~Jung Shin, Kyunghyun Sung, and
  Jong~Chul Ye.
\newblock Deep learning with domain adaptation for accelerated
  projection-reconstruction {MR}.
\newblock \emph{Magnetic resonance in medicine}, 80\penalty0 (3):\penalty0
  1189--1205, 2018.

\bibitem[Havaei et~al.(2016)Havaei, Guizard, Larochelle, and
  Jodoin]{havaei2016deep}
Mohammad Havaei, Nicolas Guizard, Hugo Larochelle, and Pierre-Marc Jodoin.
\newblock Deep learning trends for focal brain pathology segmentation in {MRI}.
\newblock In \emph{Machine Learning for Health Informatics}, pages 125--148.
  Springer, 2016.

\bibitem[He et~al.(2015)He, Zhang, Ren, and Sun]{he2015delving}
Kaiming He, Xiangyu Zhang, Shaoqing Ren, and Jian Sun.
\newblock Delving deep into rectifiers: Surpassing human-level performance on
  imagenet classification.
\newblock In \emph{Proceedings of the IEEE international conference on computer
  vision}, pages 1026--1034, 2015.

\bibitem[Hoffman et~al.(2013)Hoffman, Blei, Wang, and
  Paisley]{hoffman2013stochastic}
Matthew~D Hoffman, David~M Blei, Chong Wang, and John Paisley.
\newblock Stochastic variational inference.
\newblock \emph{The Journal of Machine Learning Research}, 14\penalty0
  (1):\penalty0 1303--1347, 2013.

\bibitem[Isensee et~al.(2018)Isensee, Petersen, Klein, Zimmerer, Jaeger, Kohl,
  Wasserthal, Koehler, Norajitra, Wirkert, et~al.]{isensee2018nnu}
Fabian Isensee, Jens Petersen, Andre Klein, David Zimmerer, Paul~F Jaeger,
  Simon Kohl, Jakob Wasserthal, Gregor Koehler, Tobias Norajitra, Sebastian
  Wirkert, et~al.
\newblock nnu-net: Self-adapting framework for u-net-based medical image
  segmentation.
\newblock \emph{arXiv preprint arXiv:1809.10486}, 2018.

\bibitem[Ivanov et~al.(2018)Ivanov, Sharaev, Artemov, Kondratyeva,
  Sushchinskaya, Burnaev, and Bernstein]{DepressionAWE2018}
S.~Ivanov, M.~Sharaev, A.~Artemov, E.~Kondratyeva, S.~Sushchinskaya,
  E.~Burnaev, and A.~Bernstein.
\newblock Learning connectivity patterns via graph kernels for fmri-based
  depression diagnostics.
\newblock In \emph{Proc. of IEEE International Conference on Data Mining
  Workshops (ICDMW)}, pages 308--314, 2018.

\bibitem[Jordan et~al.(1999)Jordan, Ghahramani, Jaakkola, and
  Saul]{jordan1999introduction}
Michael~I Jordan, Zoubin Ghahramani, Tommi~S Jaakkola, and Lawrence~K Saul.
\newblock An introduction to variational methods for graphical models.
\newblock \emph{Machine learning}, 37\penalty0 (2):\penalty0 183--233, 1999.

\bibitem[Kao et~al.(2018)Kao, Ngo, Zhang, Chen, and Manjunath]{kao_brain_2018}
Po-Yu Kao, Thuyen Ngo, Angela Zhang, Jefferson~W Chen, and BS~Manjunath.
\newblock Brain tumor segmentation and tractographic feature extraction from
  structural mr images for overall survival prediction.
\newblock \emph{arXiv preprint arXiv:1807.07716}, 2018.

\bibitem[Kingma and Welling(2014)]{kingma_vae}
Diederik~P Kingma and Max Welling.
\newblock Auto-encoding variational bayes.
\newblock \emph{stat}, 1050:\penalty0 1, 2014.

\bibitem[Kingma et~al.(2015)Kingma, Salimans, and
  Welling]{kingma_variational_2015}
Durk~P Kingma, Tim Salimans, and Max Welling.
\newblock Variational {Dropout} and the {Local} {Reparameterization} {Trick}.
\newblock \emph{Advances in Neural Information Processing Systems}, pages
  2575--2583, 2015.

\bibitem[Kohl et~al.(2017)Kohl, Bonekamp, Schlemmer, Yaqubi, Hohenfellner,
  Hadaschik, Radtke, and Maier-Hein]{kohl_adversarial_2017}
Simon Kohl, David Bonekamp, Heinz-Peter Schlemmer, Kaneschka Yaqubi, Markus
  Hohenfellner, Boris Hadaschik, Jan-Philipp Radtke, and Klaus Maier-Hein.
\newblock Adversarial networks for the detection of aggressive prostate cancer.
\newblock \emph{arXiv preprint arXiv:1702.08014}, 2017.

\bibitem[Li et~al.(2018)Li, Parikh, and He]{li2018novel}
Hailong Li, Nehal~A Parikh, and Lili He.
\newblock A novel transfer learning approach to enhance deep neural network
  classification of brain functional connectomes.
\newblock \emph{Frontiers in neuroscience}, 12:\penalty0 491, 2018.

\bibitem[Litjens et~al.(2017)Litjens, Kooi, Bejnordi, Setio, Ciompi,
  Ghafoorian, Van Der~Laak, Van~Ginneken, and S{\'a}nchez]{litjens2017survey}
Geert Litjens, Thijs Kooi, Babak~Ehteshami Bejnordi, Arnaud Arindra~Adiyoso
  Setio, Francesco Ciompi, Mohsen Ghafoorian, Jeroen~Awm Van Der~Laak, Bram
  Van~Ginneken, and Clara~I S{\'a}nchez.
\newblock A survey on deep learning in medical image analysis.
\newblock \emph{Medical image analysis}, 42:\penalty0 60--88, 2017.

\bibitem[Livne et~al.(2019)Livne, Rieger, Aydin, Taha, Akay, Kossen, Sobesky,
  Kelleher, Hildebrand, Frey, and Madai]{livne_u-net_2019}
Michelle Livne, Jana Rieger, Orhun~Utku Aydin, Abdel~Aziz Taha, Ela~Marie Akay,
  Tabea Kossen, Jan Sobesky, John~D. Kelleher, Kristian Hildebrand, Dietmar
  Frey, and Vince~Istvan Madai.
\newblock A {U}-{Net} {Deep} {Learning} {Framework} for {High} {Performance}
  {Vessel} {Segmentation} in {Patients} with {Cerebrovascular} {Disease}.
\newblock \emph{Frontiers in Neuroscience}, 13, 2019.
\newblock ISSN 1662-453X.
\newblock \doi{10.3389/fnins.2019.00097}.
\newblock URL
  \url{https://www.frontiersin.org/articles/10.3389/fnins.2019.00097/abstract}.

\bibitem[Lu(2017)]{lu2017bernstein}
Yulong Lu.
\newblock On the bernstein-von mises theorem for high dimensional nonlinear
  bayesian inverse problems.
\newblock \emph{arXiv preprint arXiv:1706.00289}, 2017.

\bibitem[Margeta et~al.(2017)Margeta, Criminisi, Cabrera~Lozoya, Lee, and
  Ayache]{margeta2017fine}
Jan Margeta, Antonio Criminisi, R~Cabrera~Lozoya, Daniel~C Lee, and Nicholas
  Ayache.
\newblock Fine-tuned convolutional neural nets for cardiac {MRI} acquisition
  plane recognition.
\newblock \emph{Computer Methods in Biomechanics and Biomedical Engineering:
  Imaging \& Visualization}, 5\penalty0 (5):\penalty0 339--349, 2017.

\bibitem[Menze et~al.(2015)Menze, Jakab, Bauer, Kalpathy-Cramer, Farahani,
  Kirby, Burren, Porz, Slotboom, Wiest, et~al.]{menze_multimodal_2015}
Bjoern~H Menze, Andras Jakab, Stefan Bauer, Jayashree Kalpathy-Cramer, Keyvan
  Farahani, Justin Kirby, Yuliya Burren, Nicole Porz, Johannes Slotboom, Roland
  Wiest, et~al.
\newblock The multimodal brain tumor image segmentation benchmark (brats).
\newblock \emph{IEEE transactions on medical imaging}, 34\penalty0
  (10):\penalty0 1993, 2015.

\bibitem[Milletari et~al.(2016)Milletari, Navab, and
  Ahmadi]{milletari_v-net:_2016}
Fausto Milletari, Nassir Navab, and Seyed-Ahmad Ahmadi.
\newblock V-net: Fully convolutional neural networks for volumetric medical
  image segmentation.
\newblock In \emph{3D Vision (3DV), 2016 Fourth International Conference on},
  pages 565--571. IEEE, 2016.

\bibitem[Mlynarski et~al.(2018)Mlynarski, Delingette, Criminisi, and
  Ayache]{mlynarski2018deep}
Pawel Mlynarski, Herv{\'e} Delingette, Antonio Criminisi, and Nicholas Ayache.
\newblock Deep learning with mixed supervision for brain tumor segmentation.
\newblock \emph{arXiv preprint arXiv:1812.04571}, 2018.

\bibitem[Myronenko(2018)]{myronenko_3d_2018}
Andriy Myronenko.
\newblock 3d {MRI} brain tumor segmentation using autoencoder regularization.
\newblock \emph{arXiv preprint arXiv:1810.11654}, 2018.

\bibitem[Neklyudov et~al.(2017)Neklyudov, Molchanov, Ashukha, and
  Vetrov]{neklyudov2017structured}
Kirill Neklyudov, Dmitry Molchanov, Arsenii Ashukha, and Dmitry~P Vetrov.
\newblock Structured bayesian pruning via log-normal multiplicative noise.
\newblock In \emph{Advances in Neural Information Processing Systems}, pages
  6775--6784, 2017.

\bibitem[Oktay et~al.(2018)Oktay, Ferrante, Kamnitsas, Heinrich, Bai,
  Caballero, Cook, de~Marvao, Dawes, O‘Regan,
  et~al.]{oktay_anatomically_2018}
Ozan Oktay, Enzo Ferrante, Konstantinos Kamnitsas, Mattias Heinrich, Wenjia
  Bai, Jose Caballero, Stuart~A Cook, Antonio de~Marvao, Timothy Dawes,
  Declan~P O‘Regan, et~al.
\newblock Anatomically constrained neural networks {(ACNNs)}: application to
  cardiac image enhancement and segmentation.
\newblock \emph{IEEE transactions on medical imaging}, 37\penalty0
  (2):\penalty0 384--395, 2018.

\bibitem[Pan and Yang(2010)]{pan2010survey}
Sinno~Jialin Pan and Qiang Yang.
\newblock A survey on transfer learning.
\newblock \emph{IEEE Transactions on knowledge and data engineering},
  22\penalty0 (10):\penalty0 1345--1359, 2010.

\bibitem[Pominova et~al.(2018)Pominova, Artemov, Sharaev, Kondrateva, Cichocki,
  Burnaev, and Bernstein]{3DANNMRI2018}
M.~Pominova, A.~Artemov, M.~Sharaev, E.~Kondrateva, A.~Cichocki, E.~Burnaev,
  and A.~Bernstein.
\newblock Voxelwise 3d convolutional and recurrent neural networks for epilepsy
  and depression diagnostics from structural and functional mri data.
\newblock In \emph{Proc. of IEEE International Conference on Data Mining
  Workshops (ICDMW)}, pages 299--307, 2018.

\bibitem[Rey et~al.(2002)Rey, Subsol, Delingette, and
  Ayache]{rey_automatic_2002}
David Rey, Gérard Subsol, Hervé Delingette, and Nicholas Ayache.
\newblock Automatic detection and segmentation of evolving processes in 3d
  medical images: {Application} to multiple sclerosis.
\newblock \emph{Medical Image Analysis}, 6\penalty0 (2):\penalty0 163--179,
  June 2002.
\newblock ISSN 1361-8415.
\newblock \doi{10.1016/S1361-8415(02)00056-7}.
\newblock URL
  \url{http://www.sciencedirect.com/science/article/pii/S1361841502000567}.

\bibitem[Ronneberger et~al.(2015)Ronneberger, Fischer, and
  Brox]{ronneberger2015u}
Olaf Ronneberger, Philipp Fischer, and Thomas Brox.
\newblock U-net: Convolutional networks for biomedical image segmentation.
\newblock In \emph{International Conference on Medical image computing and
  computer-assisted intervention}, pages 234--241. Springer, 2015.

\bibitem[Shah et~al.(2018)Shah, Merchant, and Awate]{shah2018ms}
Meet~P Shah, SN~Merchant, and Suyash~P Awate.
\newblock Ms-net: Mixed-supervision fully-convolutional networks for
  full-resolution segmentation.
\newblock In \emph{International Conference on Medical Image Computing and
  Computer-Assisted Intervention}, pages 379--387. Springer, 2018.

\bibitem[Sharaev et~al.(2018{\natexlab{a}})Sharaev, Andreev, Artemov, Burnaev,
  Kondratyeva, Sushchinskaya, Samotaeva, Gaskin, and Bernstein]{Pipeline2018}
M.~Sharaev, A.~Andreev, A.~Artemov, E.~Burnaev, E.~Kondratyeva,
  S.~Sushchinskaya, I.~Samotaeva, V.~Gaskin, and A.~Bernstein.
\newblock Pattern recognition pipeline for neuroimaging data.
\newblock In Luca Pancioni, Friedhelm Schwenker, and Edmondo Trentin, editors,
  \emph{Artificial Neural Networks in Pattern Recognition}, pages 306--319,
  Cham, 2018{\natexlab{a}}. Springer International Publishing.
\newblock ISBN 978-3-319-99978-4.

\bibitem[Sharaev et~al.(2018{\natexlab{b}})Sharaev, Artemov, Kondratyeva,
  Sushchinskaya, Burnaev, Bernstein, Akzhigitov, and Andreev]{Epilepsy2018}
M.~Sharaev, A.~Artemov, E.~Kondratyeva, S.~Sushchinskaya, E.~Burnaev,
  A.~Bernstein, R.~Akzhigitov, and A.~Andreev.
\newblock Mri-based diagnostics of depression concomitant with epilepsy: in
  search of the potential biomarkers.
\newblock In \emph{Proceedings of IEEE 5th International Conference on Data
  Science and Advanced Analytics}, pages 555--564, 2018{\natexlab{b}}.

\bibitem[Sheline(2000)]{sheline20003d}
Yvette~I Sheline.
\newblock 3d {MRI} studies of neuroanatomic changes in unipolar major
  depression: the role of stress and medical comorbidity.
\newblock \emph{Biological psychiatry}, 48\penalty0 (8):\penalty0 791--800,
  2000.

\bibitem[Simpson et~al.(2019)Simpson, Antonelli, Bakas, Bilello, Farahani, van
  Ginneken, Kopp-Schneider, Landman, Litjens, Menze, Ronneberger, Summers,
  Bilic, Christ, Do, Gollub, Golia-Pernicka, Heckers, Jarnagin, McHugo, Napel,
  Vorontsov, Maier-Hein, and Cardoso]{simpson2019large}
Amber~L. Simpson, Michela Antonelli, Spyridon Bakas, Michel Bilello, Keyvan
  Farahani, Bram van Ginneken, Annette Kopp-Schneider, Bennett~A. Landman,
  Geert Litjens, Bjoern Menze, Olaf Ronneberger, Ronald~M. Summers, Patrick
  Bilic, Patrick~F. Christ, Richard K.~G. Do, Marc Gollub, Jennifer
  Golia-Pernicka, Stephan~H. Heckers, William~R. Jarnagin, Maureen~K. McHugo,
  Sandy Napel, Eugene Vorontsov, Lena Maier-Hein, and M.~Jorge Cardoso.
\newblock A large annotated medical image dataset for the development and
  evaluation of segmentation algorithms, 2019.

\bibitem[Smith(2017)]{smith2017cyclical}
Leslie~N Smith.
\newblock Cyclical learning rates for training neural networks.
\newblock In \emph{2017 IEEE Winter Conference on Applications of Computer
  Vision (WACV)}, pages 464--472. IEEE, 2017.

\bibitem[Van~Opbroek et~al.(2015)Van~Opbroek, Ikram, Vernooij, and
  De~Bruijne]{van2015transfer}
Annegreet Van~Opbroek, M~Arfan Ikram, Meike~W Vernooij, and Marleen De~Bruijne.
\newblock Transfer learning improves supervised image segmentation across
  imaging protocols.
\newblock \emph{IEEE transactions on medical imaging}, 34\penalty0
  (5):\penalty0 1018--1030, 2015.

\bibitem[Wachinger et~al.(2018)Wachinger, Reuter, and
  Klein]{wachinger_deepnat:_2018}
Christian Wachinger, Martin Reuter, and Tassilo Klein.
\newblock {DeepNAT}: {Deep} convolutional neural network for segmenting
  neuroanatomy.
\newblock \emph{NeuroImage}, 170:\penalty0 434--445, April 2018.
\newblock ISSN 1053-8119.
\newblock \doi{10.1016/j.neuroimage.2017.02.035}.
\newblock URL
  \url{http://www.sciencedirect.com/science/article/pii/S1053811917301465}.

\bibitem[Wilson and Cook(2018)]{wilson2018adversarial}
Garrett Wilson and Diane~J Cook.
\newblock Adversarial transfer learning.
\newblock \emph{arXiv preprint arXiv:1812.02849}, 2018.

\bibitem[Zhou et~al.(2017)Zhou, Shin, Zhang, Gurudu, Gotway, and
  Liang]{zhou2017fine}
Zongwei Zhou, Jae Shin, Lei Zhang, Suryakanth Gurudu, Michael Gotway, and
  Jianming Liang.
\newblock Fine-tuning convolutional neural networks for biomedical image
  analysis: actively and incrementally.
\newblock In \emph{Proceedings of the IEEE conference on computer vision and
  pattern recognition}, pages 7340--7351, 2017.

\end{thebibliography}

\newpage
\section*{A. Stohastic Variational Inference} \label{sec:deep_bayes}

Variational inference \citep{jordan1999introduction} introduces approximate posterior distribution $q_\theta(w)$ from some parametric family, e.g. fully factorized Gaussian, and solve optimization problem, minimizing Kullback–Leibler divergence between true posterior distribution $p(w|\mathcal{D})$ and variational approximation $q_\theta(w)$ with respect to parameters $\theta$.

\begin{equation}\label{eq:vi-trg}
\min \text{KL}\left(q_\theta(w)) || p(w|\mathcal{D})\right).
\end{equation}

Where Kullback–Leibler divergence, or KL-divergence is defined as:

\begin{equation*}
\text{KL}\left(q(x)) || p(x)\right) = -\int q(x) \log \left(\frac{p(x)}{q(x)} \right).
\end{equation*}

Note that equation \eqref{eq:vi-trg} still contains posterior distribution, which is not known. Let us rewrite this equation in the following way:

\begin{align*}
\text{KL}\left(q_\theta(w)) || p(w|\mathcal{D})\right) &= \mathbb{E}_{q_\theta(w)} \log \frac{q_\theta(w) p(\mathcal{D})}{p(\mathcal{D}|w)p(w)} = \\
& =\log p(\mathcal{D}) + \mathbb{E}_{q_\theta(w)}\log \frac{q_\theta(w)}{p(w)} - \mathbb{E}_{q_\theta(w)}\log p(\mathcal{D}|w) = \\
&= \log p(\mathcal{D}) - \mathcal{L}(\theta).
\end{align*}

Above we have received a decomposition of the marginal log-likelihood into two components: the first one is KL-divergence between exact posterior and its variational approximation, while the second one is so-called evidence lower bound (ELBO, $\mathcal{L}(\theta)$).

\begin{equation*}
    \max_{\theta} \log p(\mathcal{D}) = \max_{\theta}\left[ \text{KL}\left(q_\theta(w)) || p(w|\mathcal{D})\right) + \mathcal{L}(\theta) \right].
\end{equation*}

If variational posterior is precisely equal to the true posterior, KL-divergence is zero and ELBO coincides with marginal log-likelihood. Since KL-divergence is always non-negative, ELBO cannot be greater than $\log p(\mathcal{D})$ and thus problem reduces to ELBO maximization.

\begin{equation*}
  \mathcal{L}(\theta) =  \mathbb{E}_{q_{\theta}(w)}\log p(\mathcal{D}|w) - \mathbb{E}_{q_\theta(w)}\log \frac{q_\theta(w)}{p(w)}= \mathcal{L_D} - \text{KL}\left(q_{\theta}(w)||p(w)\right).
\end{equation*}

 The first part of the target function is data term $\mathcal{L_{D}}$ also referred to as a reconstruction error. It is in charge of prediction quality, forcing the model to fit the data. Second term --- Kullback–Leibler divergence between a variational distribution and prior over parameters of the model requires posterior distribution to be as close as possible to the prior, serving among other things as a regularization.

In complex models, such as neural networks, it is not trivial to compute gradients of the data term $ \mathcal{L_D}$. In practice one may overcome this difficulty with the help of sampling and reparametrization trick, resulting in so-called doubly stochastic variational inference \citep{kingma_variational_2015}. Let $\mathcal{D}^*$ be minibatch of size $M < N$ and  $w = f(\theta, \varepsilon_i)$ a representation of the parametric random variable $w \sim q_\theta(w)$ as a deterministic function of the non-parametric noise $\varepsilon \sim p(\varepsilon)$ . Then the unbiased Monte Carlo estimate of the data term $\mathcal{L_{D}}$ has the following form:

\begin{equation*}
\mathcal{L_{D}} \approx \mathcal{L_{D^*}} = \frac{N}{M}\sum_{i = 1}^M \log p(\mathcal{D}_i|f(\theta, \widehat{\varepsilon}_i)), \quad \widehat{\varepsilon}_i  \sim p(\varepsilon).
\end{equation*}

We apply doubly stochastic variational inference framework\citep{kingma_variational_2015} to the U-net model. Dataset in this case contains pairs of images $\{x_i\}_{i=1}^N$ and their masks $\{y_i\}_{i=1}^N$. All the parameters of the model are of the form of convolutional filters $(w^{(1)}, \ldots, w^{(L)})$, where $L$ is the number of convolutional layers. We assume that both variational approximation $q_{\theta}(w)$ and prior distribution $p(w)$ are factorized over layers, input and output channels:

\begin{align*}
    q_{\theta}(w) &= \prod_{i = 1}^L \prod_{p = 1}^{C_{inp}^{(i)}} \prod_{k = 1}^{C_{out}^{(i)}} q_{\theta_{i p k}}(w^{(i)}_{p, k}),\\
    p(w) &= \prod_{i = 1}^L \prod_{p = 1}^{C_{inp}^{(i)}} \prod_{k = 1}^{C_{out}^{(i)}} p(w^{(i)}_{p, k}).
\end{align*}
where $C_{inp}^{(i)}, C_{out}^{(i)}$ --- the number of input and output channels on the i-th layer of the network. 

Taking into account both reparametrization trick and factorization of the distributions, the final optimization task is the follwing:

\begin{equation*}
\max_{\theta} \mathcal{L}(\theta) \approx  \max_{\theta} \mathcal{L_{D^*}} - \sum_{i, p, k} \text{KL}\left(q_{\theta_{i p k}}(w^{(i)}_{p, k}||p(w^{(i)}_{p, k})\right).
\end{equation*}

\newpage
\section*{B. Architecture details} \label{sec:architectures}
\subsection*{3D U-Net}
\small{
\begin{verbatim}
ConvBlock(in_channels, out_channels, s) = 
    Sequential(
        (0): InstanceNorm3d(in_channels)
        (1): ReLU()
        (2): Conv3d(in_channels, out_channels, kernel_size=(3, 3, 3), stride=(s, s, s)))

  UNet3D(
  (init_conv): Conv3d(1, 16, kernel_size=(3, 3, 3), stride=(1, 1, 1))
  (down1): BasicDownBlock(
    (conv_1): ConvBlock(16, 32, 2)
    (conv_2): ConvBlock(32, 32, 1)
    (down): ConvBlock(16, 32, 2)
    )
  (down2): BasicDownBlock(
    (conv_1): ConvBlock(32, 32, 1)
    (conv_2): ConvBlock(32, 32, 1)
  )
  (down3): BasicDownBlock(
    (conv_1): ConvBlock(32, 32, 2)
    (conv_2): ConvBlock(32, 32, 1)
    (down): ConvBlock(32, 32, 2)
  )
  (down4): BasicDownBlock(
    (conv_1): ConvBlock(32, 32, 1)
    (conv_2): ConvBlock(32, 32, 1)
  )
  (down5): BasicDownBlock(
    (conv_1): ConvBlock(32, 64, 2)
    (conv_2): ConvBlock(64, 64, 1)
    (down): ConvBlock(32, 64, 2)
  (down6): BasicDownBlock(
    (conv_1): ConvBlock(64, 64, 1)
    (conv_2): ConvBlock(64, 64, 1)
  )
  (up1): BasicUpBlock(
    (upsample): Sequential(
      (0): ConvBlock(64, 32, 1)
      (1): Upsample(scale_factor=2.0, mode=trilinear)
    )
    (conv_1): ConvBlock(32, 32, 1)
    (conv_2): ConvBlock(32, 32, 1)
  )
  (up2): BasicUpBlock(
    (upsample): Sequential(
      (0): ConvBlock(32, 32, 1)
      (1): Upsample(scale_factor=2.0, mode=trilinear)
    )
    (conv_1): ConvBlock(32, 32, 1)
    (conv_2): ConvBlock(32, 32, 1)
  )
  (up3): BasicUpBlock(
    (upsample): Sequential(
      (0): ConvBlock(32, 16, 1)
      (1): Upsample(scale_factor=2.0, mode=trilinear)
    )
    (conv_1): ConvBlock(16, 16, 1)
    (conv_2): ConvBlock(16, 16, 1)
  )
  (out): Conv3d(16, 2, kernel_size=(1, 1, 1), stride=(1, 1, 1))
)
\end{verbatim}}

\subsection*{VAE for DWP}
\small{
\begin{verbatim}
Kernel_3D_VAE(
  (encode): Sequential(
    (0): Conv3d(1, 32, kernel_size=(3, 3, 3), stride=(1, 1, 1))
    (1): MaxPool3d(kernel_size=2)
    (2): ELU(alpha=1.0)
    (3): Conv3d(32, 64, kernel_size=(3, 3, 3), stride=(1, 1, 1))
    (4): MaxPool3d(kernel_size=2)
    (5): ELU(alpha=1.0)
    (6): Conv3d(64, 128, kernel_size=(1, 1, 1), stride=(1, 1, 1))
    (7): ELU(alpha=1.0)
    (8): Flatten()
  )
  (latent_mu): Linear(in_features=128, out_features=6)
  (latent_logsigma): Linear(in_features=128, out_features=6)
  (linear): Linear(in_features=6, out_features=128)
  (decode): Sequential(
    (0): Conv3d(128, 128, kernel_size=(3, 3, 3), stride=(1, 1, 1))
    (1): ELU(alpha=1.0)
    (2): ConvTranspose3d(128, 128, kernel_size=(3, 3, 3), stride=(1, 1, 1))
    (3): ELU(alpha=1.0)
    (4): ConvTranspose3d(128, 64, kernel_size=(1, 1, 1), stride=(1, 1, 1))
    (5): ELU(alpha=1.0)
    (6): ConvTranspose3d(64, 32, kernel_size=(1, 1, 1), stride=(1, 1, 1))
    (7): ELU(alpha=1.0)
  )
  (reconstruction_mu): Sequential(
    (0): ConvTranspose3d(32, 1, kernel_size=(1, 1, 1), stride=(1, 1, 1))
    (1): Tanh()
  )
  (reconstruction_logsigma): Sequential(
    (0): ConvTranspose3d(32, 1, kernel_size=(1, 1, 1), stride=(1, 1, 1))
    (1): Tanh()
  ))
\end{verbatim}}

\newpage
\section*{C. Example predictions} \label{sec:prediction}

\end{document}